\documentclass[preprint,12pt]{elsarticle}

\usepackage{graphicx}
\usepackage{booktabs,caption}
\usepackage[flushleft]{threeparttable}
\usepackage{bm}
\usepackage{newtxtext,newtxmath}
\usepackage{hyperref}
\usepackage{comment}
\usepackage[T1]{fontenc}
\usepackage[utf8]{inputenc}

\hypersetup{
    colorlinks=true,
    citecolor=blue,
    linkcolor=blue,
    filecolor=magenta,      
    urlcolor=cyan}
\usepackage{lipsum}

\journal{Physics of the Dark Universe}

\begin{document}

\begin{frontmatter}

\title{Quasar cosmology II: joint analyses with Cosmic Microwave Background}

\author[1,2]{M. Benetti}
\author[1,2]{G. Bargiacchi}
\author[3,4]{G. Risaliti}
\author[1,2,5]{S. Capozziello}
\author[3,4]{E. Lusso}
\author[4,6]{M. Signorini}

\affiliation[1]{organization={Scuola Superiore Meridionale},
            addressline={Via Mezzocannone 4}, 
            city={Napoli},
            postcode={80134}, 
            country={Italy}}
            
\affiliation[2]{organization={Istituto Nazionale di Fisica Nucleare (INFN), Laboratori Nazionali di Frascati (LNF)},
            addressline={Via E. Fermi 54}, 
            city={Frascati, Roma},
            postcode={00044}, 
            country={Italy}}

\affiliation[3]{organization={ Dipartimento di Fisica e Astronomia, Universit\`{a} degli Studi di Firenze},
        addressline={via G. Sansone 1},
        city={Sesto Fiorentino, Firenze},
        postcode={50019}, 
        country={Italy}}

\affiliation[4]{organization={Istituto Nazionale di Astrofisica (INAF) Osservatorio Astrofisico di Arcetri},
        addressline={Largo Enrico Fermi 5},
        city={Firenze},
        postcode={I-50125}, 
        country={Italy}}      
            
\affiliation[5]{organization={Dipartimento di Fisica  ``E. Pancini", Universit\`{a} di Napoli  ``Federico II"},
            addressline={Via Cinthia 9}, 
            city={Napoli},
            postcode={I-80126}, 
            country={Italy}}

 \affiliation[6]{organization={Dipartimento di Matematica e Fisica, Universit\`{a} di Roma 3},
            addressline={Via della Vasca Navale 84}, 
            city={Roma},
            postcode={00146}, 
            country={Italy}}

   \date{Received XX; accepted XX}
   
\begin{abstract}

Currently, the increasing availability of accurate cosmological probes leads to the emergence of tensions between data on the one hand and between theoretical predictions and direct observations on the other. Moreover, after 25 years since the discovery of the accelerated expansion of the Universe has elected the $\Lambda$CDM model as the reference model, resolving shortcomings of the standard cosmological model seems to be an unpostponed priority. 
Hence, it is key to test alternative models and investigate new cosmological probes at distances that range from the late to the early Universe, namely between the cosmic microwave background (CMB) and type Ia supernovae and baryonic acoustic oscillations (BAO) data.
Bargiacchi et al. (2022) for the first time analysed dark energy (DE) models using quasars (QSOs) while also testing their consistency with BAO. Here, we carry on by exploring the compatibility of QSOs with both CMB data and dark energy survey measurements against the standard cosmological model and some DE extensions, such as the $w$CDM and Chevallier-Polarski-Linder parameterisations. We also consider an interacting dark matter and vacuum energy scenario, where vacuum energy perturbations affect the evolution of the matter growth rate in a decomposed Chaplygin gas model. 
We implement the QSO probe in Cobaya Markov chain Monte Carlo algorithm, using Botzmann solver codes as Cosmic Linear Anisotropy Solving System (CLASS) for the theory predictions. 
Our work shows that simple DE deviations from $\Lambda$CDM model do not reconcile the data and that only more complex models of interaction in the dark sector can succeed in solving the discrepancies of probes at all scales. 
\end{abstract}


\end{frontmatter}


\section{Introduction}
\label{sec:intro}

The spatially flat $\Lambda$ cold dark matter (CDM) has been considered by several decades the cosmological model that best describes the Universe and reproduces the current observations. Indeed, this model intrinsically predicts the current acceleration of the Universe's expansion discovered from type Ia supernovae \citep[SNe Ia; e.g.][]{riess1998,perlmutter1999,Pan-STARRS1:2017jku,2022ApJ...934L...7R} and the measurements from cosmic microwave background \citep[CMB; e.g.][]{planck2018,WMAP:2012nax} and baryonic acoustic oscillations \citep[BAO; e.g.][]{eboss2021,2018MNRAS.473.4773A,2011MNRAS.416.3017B,2015MNRAS.449..835R,2017MNRAS.470.2617A}. Actually, in this model the evolution of the Universe is driven by a CDM component and a dark energy (DE) which is a cosmological constant $\Lambda$ \citep{2001LRR.....4....1C}, and thus it is characterized by an equation of state (EoS) $w(z) = P_{\Lambda}/\rho_{\Lambda}=-1$, with $z$ the redshift, $P_{\Lambda}$ and $\rho_{\Lambda}$ the pressure and energy density of DE, respectively. 

Beside the success of the $\Lambda$CDM model, it suffers from longstanding issues: the cosmological constant problem \citep{1989RvMP...61....1W}, the fine-tuning problem \citep{deCarlos:1993rbr}, and the unknown physical ground of CDM and DE \citep{Gaskins:2016cha,Frusciante:2019xia}. 
In addition, also the assumption of a null curvature of the Universe has been questioned \citep{Park:2017xbl,2020NatAs...4..196D,Handley:2019tkm,DiValentino:2020hov,2021JCAP...10..008Y,Dhawan:2021mel,vagnozzi:2020dfn,Vagnozzi:2020rcz,Abdalla:2022yfr}, along with the actual evolution of the DE \citep{DiValentino:2021izs,Benetti:2019lxu,2021JCAP...10..008Y,Abdalla:2022yfr, Salzano:2021zxk}. Furthermore, a deeper inspection of the $\Lambda$CDM model and the search for cosmological models alternative to the standard one have been recently boosted by the arising of the Hubble constant, $H_0$, tension \citep[see][for a review]{2022JHEAp..34...49A}. Indeed, the value of $H_0$ measured in the late Universe from SNe Ia and Cepheids ($H_0 = 73.04 \pm 1.04  \, \mathrm{km} \, \mathrm{s}^{-1} \, \mathrm{Mpc}^{-1}$, \citep{2022ApJ...934L...7R}) and the one derived from the Planck data of the CMB by assuming a flat $\Lambda$CDM model ($H_0 = 67.4 \pm 0.5  \, \mathrm{km} \, \mathrm{s}^{-1} \, \mathrm{Mpc}^{-1}$, \citep{planck2018}) are significantly inconsistent, with a discrepancy that ranges from 4 to 6 $\sigma$ deviation depending on the samples investigated \citep{2019ApJ...876...85R,2020PhRvR...2a3028C,2020MNRAS.498.1420W}. This problem proves to be exacerbated when this tension is investigated with other cosmological probes, e.g the $H_0$ value of the CMB is favored by cosmic chronometers \citep{2018JCAP...04..051G,Favale:2023lnp,vagnozzi:2020dfn}, that cover a redshift range up to $z \sim 2$ and do not assume any a-priori cosmological model, but time-delay and strong lensing of quasars (QSOs) between $z=0.6$ and $z=1.8$ point toward the $H_0$ measured with SNe Ia \citep{2019ApJ...886L..23L}, while some other probes, such as QSOs \citep{2023ApJS..264...46L}, the tip of the red-giant branch (TRGB; \citep{2021ApJ...919...16F}), and gamma-ray bursts (GRBs; \citep{2009MNRAS.400..775C, postnikov14,2022Galax..10...24D,2022MNRAS.tmp.2639D,2022PASJDainotti}), indicate a $H_0$ value intermediate between the two. In this framework, different approaches have been attempted both to investigate the reliability of the flat $\Lambda$CDM model and solve its shortcomings. These efforts range from cosmographic analyses \citep{Aviles,Chebyshev,2017A&A...598A.113D,2020JCAP...12..007Z,lusso2019,rl19,2021A&A...649A..65B,Benetti} to additional relativistic particles, time-dependent DE EoS, modified theories of gravity, and possible evolution of $H_0$ \citep[e.g.][]{RoccoReview,Spallicci,2020ApJ...894...54D,2020arXiv201213932A,2020PhRvD.101l3516A,2020arXiv201110559R,Dainotti2021ApJ...912..150D,2022Galax..10...24D,2023MNRAS.522L..72S,Capozziello:2023ewq}. 

A crucial point in this scenario is the gap of information between the furthest SNe Ia observation, at $z=2.26$ \citep{Rodney}, and the CMB, giving information at $z \sim 1100$. For this reason, the cosmological community is striving to find new cosmological sources that can probe the Universe at intermediate epochs. In this regard, QSOs 
, observed 
up to $z=7.64$ \citep{2021ApJ...907L...1W} 
, have been recently turned into cosmological tools. Indeed, QSOs are employed in cosmological studies through the non-linear relation between their ultraviolet (UV) and X-ray luminosities \citep[e.g.][]{steffen06,just07,2010A&A...512A..34L,lr16,2020A&A...642A.150L,2021arXiv210903252B} and this relation has been applied in several analyses in the QSO realm \citep{2014JCAP...01..027M,rl19,lusso19,2021A&A...649A..65B,2022MNRAS.515.1795B,2023ApJ...950...45D,2023ApJS..264...46L}. 
Since QSOs 
can probe the Universe between SNe Ia and the CMB, they have also been used jointly, and in combination with other probes, to investigate cosmological models \citep{lusso19,2023MNRAS.521.3909B,2023ApJ...951...63D}.

Concerning the combination of different data sets, the importance of checking the consistency among the individual probes has recently been pinpointed \citep{2021ApJ...908...84V,2021JCAP...11..060G,2022MNRAS.515.1795B,2023ApJ...951...63D}. Indeed, a joint analysis is physically and statistically validated only if all the probes are compatible in the multi-dimensional space of the free parameters of the investigated model. Given the significance of this latter point, we start our work exactly by testing the consistency of the data sets we use. More specifically, as a natural extension of \citep{2022MNRAS.515.1795B}, we consider five probes, SNe Ia, QSOs, BAO, CMB, and dark energy survey (DES), and six cosmological models, which are the flat and non-flat $\Lambda$CDM, the flat and non-flat $w$CDM, the Chevallier-Polarski-Linder (CPL) parameterisation, and an interacting dark energy (iDE) model of generalized Chaplygin gas \citep{Benetti:2019lxu, Benetti:2021div,Borges:2023xwx,Borges:2013bya,Carneiro:2014jza}. Thus, we fit each of these models with all the probes separately and join the different data sets in case they proved to be consistent in the parameter space. This analysis allows us to compare the constraints on the free parameters provided by sources on different scales and on a wide range of distances to search for a cosmological model that solves the tensions among different probes.

To perform our analysis, we implement the QSO data in Monte Carlo Markov Chains (MCMC) algorithms such as Cobaya \citep{Torrado:2020dgo}, this allows us to combine, for the first time, QSOs with full likelihood of \textit{Planck} and DES. Furthermore, the comparison between QSOs, SNe Ia, and BAO enables us to validate the results presented in \citep{2022MNRAS.515.1795B} by employing a different and independent methodology. In addition, we investigate the consistencies or discrepancies among the investigated probes to validate a joint analysis. Furthermore, we discuss in detail the impact of the initial conditions of the analysis, as the choice of the free and fixed parameters, as well as the priors, which proves to be crucial to obtain reliable and unbiased results. Our main goal is to search for a cosmological model that could remove the tensions among the different probes thus describing the physics of the Universe and its evolution starting from today and going back in time up to earliest stages of the CMB.  

The manuscript is organized as follows. Section \ref{sec:Data_sets} describes the data sets considered and details the method developed to include QSOs in MCMC algorithms. Section \ref{sec:theory} introduces the cosmological model investigated in this work, while Sect. \ref{sec:Cosmological results} provides the methodology and the main results of the analyses. In Section \ref{Conclusions} we draw our conclusions. We also focus on the impact of the analysis assumptions on the results in \ref{appendix:prior}.

\section{Data sets}
\label{sec:Data_sets}

In this paper, we consider both large scale structure data by SNe Ia, QSOs, BAO and DES experiment, and CMB. The likelihoods for these data are implemented in the publicly available MCMC algorithm we use for our analysis, Cobaya \citep{Torrado:2020dgo}, with the exception of the QSO data. 

In this section, we therefore detail the method we have developed to include the QSO analysis in these algorithms, along with the description of the different data sets considered. 

\subsection{Cosmic Microwave Background}
\label{sec:CMB}
We analyse CMB measurements, through the Planck (2018) data~\citep{Planck:2019nip}, using ``TT,TE,EE+lowE" data by combination of temperature power spectra (TT) and cross-correlation TE and EE polarization spectra over the multipole range $\ell \in [30, 2508]$, the
low-$\ell$ temperature Commander likelihood, and the low-$\ell$ SimAll EE likelihood \citep[see][for details]{Planck:2019nip}. 
Additionally, in our analysis we consider the lensing reconstruction power spectrum from the latest Planck satellite data release (2018, \citep{Planck:2019nip, Planck:2018lbu}). 
Hereafter, for sake of simplicity we refer to this data set as ``CMB".

\subsection{Supernovae Ia}
\label{sec:SNe Ia}

We use the collection of 1048 SNe Ia from the \textit{Pantheon} survey \citep{Pan-STARRS1:2017jku}  that covers the redshift range $z = 0.01 - 2.26$.  This is a very large sample, combining the subset of 279 Pan-STARRS1 SNe Ia with useful distance estimates from SuperNova Legacy Survey (SNLS), Sloan Digital Sky Survey (SDSS), low-$z$ and Hubble Space Telescope (HST) samples. 
In our analysis, we include both statistical and systematic uncertainties, also marginalising over the nuisance parameter $\mathcal{M}$ since the theoretical modelled apparent magnitude is calculated as $m_{\rm th} = \mathrm{DM}_{\rm th} + M = 5\,\log_{10} D_\mathrm{L} + \mathcal{M}$, where $\mathrm{DM}_{\rm th}$ is the theoretical distance modulus, $M$ the observed absolute magnitude and $D_\mathrm{L}$ the luminosity distance. $\mathcal{M}$ depends on $H_0$ and the absolute magnitude $M$ and for the Pantheon sample it is estimated to be $\mathcal{M} = -19.263 \pm 0.049$ \citep{Riess:2020fzl}. 
 Hereafter, we indicate this data set with ``Pth".

\subsection{Quasars}
\label{sec:qso}

As anticipated, we employ QSOs as cosmological probes by considering the non-linear relation between the nuclear UV and X-ray luminosity \citep[e.g.][]{steffen06,just07,2010A&A...512A..34L,lr16,2021arXiv210903252B}, that allows us to determine their distance making use of
\begin{equation} \label{dlqso}
\text{log}D_\mathrm{{L}}(z) = \frac{\left[\text{log}F_{\rm X} - \beta -\gamma\, (\text{log}F_{\rm UV}+27.5)\right]}{2(\gamma -1)} - \frac{1}{2}\text{log}( 4 \pi) +28.5,
\end{equation}
which is obtained from the luminosity X$-$UV relation by converting the luminosities into fluxes.
In this expression, the luminosity distance $D_\mathrm{{L}}$ is expressed in units of cm and it is normalised to 28.5 in logarithm, $F_{\rm X}$ and $F_{\rm UV}$ are the measured flux densities (in $\mathrm{erg \, s^{-1} \, cm^{-2} \, Hz^{-1}}$) at the rest-frame 2 keV and 2500 \AA, respectively, $F_{\rm UV}$ is normalised to the logarithmic value of $-$27.5, and $\gamma$ and $\beta$ are the slope and the intercept of the logarithmic X$-$UV luminosity relation, respectively. 

{A full discussion on how to employ the above relation for cosmological fits is presented in \citep{2025A&A...697A.108L}. Here we only summarize the details relevant for the present work. In particular, here we fix the $\gamma$ and $\beta$ parameters, then we derive the luminosity distances from the above equation, and we fit the so-derived Hubble diagram. In order to correctly perform this procedure, the following steps are necessary:\\
1. Determination of $\gamma$ and $\beta$. In principle, both these parameters can be left free to vary in a cosmological fit. However, iIf we marginalize over $\gamma$ and $\beta$, the presence of a tension between the observational data and the investigated cosmological model  could be alleviated or completely removed by changing the best-fit values evaluated for these parameters. More precisely, it could happen that the cosmological model is not in agreement with the probes, but it seems to well reproduce  the data, and thus that there is no discrepancy, since the tension is shifted to the values of $\gamma$ and $\beta$, which are altered compared to their intrinsic values. This outcome would be incorrect because the values of $\gamma$ and $\beta$ are not arbitrary: $\gamma$ can be measured by fitting the X-ray to UV relation in small redshift bins (this analysis is cosmology-independent provided that the redshift bins -hence the distance intervals in each bin- are sufficiently narrow); $\beta$ represents the absolute normalization of the X-ray to UV relation, and is fixed by the requirement of an optimal match between the Hubble diagram of quasars and supernovae in the common redshift interval\footnote{The need for fitting SNe Ia and QSOs jointly has already been explained in \citep{rl19}, \citep{2020A&A...642A.150L} and \citep{2022MNRAS.515.1795B} and will be addressed later in this section.} 
In order to obtain the values of $\gamma$ and $\beta$, we fitted the quasars and SNe Ia Pth sample with a cosmographic orthogonal fifth-order logarithmic polynomial described in \citep{2021A&A...649A..65B}. This procedure is completely independent from a cosmological model since it uses a cosmographic approach. In addition, it marginalises over the free parameters $\gamma$, $\beta$, and the intrinsic dispersion of the X$-$UV flux relation, $\delta$. It also applies a 3$-\sigma$ clipping selection that removes extreme outliers from the initial 2036 sources of \citep{2020A&A...642A.150L}\footnote{We here refer to the QSO sample released for cosmological applications by \citep{2020A&A...642A.150L} with the cutting at redshift $z > 0.7$ for the sources with a photometric determination of the UV flux, as explained in the same paper.} retaining 2014 QSOs.
Thus, by fixing $\gamma$ and $\beta$ to the values obtained with this cosmographic method, we avoid any dependence on an assumed cosmological model that could bias our analysis and affect our results.\\
2. Determination of the distance moduli DM. 
At this point, we can insert the obtained best-fit values of $\gamma$ and $\beta$ in Eq. \eqref{dlqso} to compute $D_{\rm L}$ for the resulting QSO sample of 2014 sources. The best-fit values given by the above-described analysis are: $\gamma = 0.591 \pm 0.011$, $\beta = -31.475 \pm 0.008$, and $\delta = 0.209 \pm 0.004$. Then, we can compute the corresponding distance moduli and their associated uncertainties $\Delta(\mathrm{DM}) = \sqrt{\sigma^{2}_{\rm DM} + \delta^{2}_{\rm DM}}$, where $\sigma_{\rm DM}$ is the statistical error propagated from $D_{\rm L}$ on the distance modulus and $\delta_{\rm DM}$ is the intrinsic dispersion of the X-UV distance modulus-redshift relation and is related to $\delta$ through $\delta_{\rm DM} = 5 \, \delta / (2 |\gamma -1|)$, with $\gamma$ and $\delta$ fixed to their best-fit values. Following these steps, we end up with distance moduli and associated uncertainties for each QSO source, as we do not have to consider any systematic error; this is exactly the QSO physical quantity we implement as input in the MCMC algorithms. Though the parameters of the X-UV relation are fixed in this procedure, we have to include and marginalise over a calibration nuisance parameter "$k$", just as required by the Pantheon code for SNe Ia in MCMC codes.\\
3. Checks of the method with different fitting procedures.}
The goodness of our novel procedure to include QSOs in cosmological analysis has been checked by comparing the Cobaya analysis results with the ones obtained with the Python package emcee \citep{2013PASP..125..306F}, which is a pure-Python implementation of Goodman \& Weare’s affine invariant MCMC ensemble sampler.
The weak differences detected between the two types of analysis are due to the fact that the latter package 
is not implemented with a Boltzmann solver code, as the first ones. 
For example, it does not include cosmological free parameters such as $\Omega_{\rm b}\, h^{2}$ and $\Omega_{\rm cdm}\, h^{2}$ (now accounted both at background and perturbation level) where $\Omega_{\rm b}$ and $\Omega_{\rm cdm}$ are the density parameters of the baryon and CDM components, respectively, and $h$ is the dimensionless Hubble constant $\displaystyle h= {H_{0}}/{100 \, \mathrm{km \,s^{-1} \, Mpc^{-1}}}$. 
This is responsible for the small differences between the results of this work and the ones reported in \citep{ 2022MNRAS.515.1795B} regarding the Pth+QSO data set, although the results are completely statistically consistent.
{For example, for a flat $w$CDM model, we obtain 1.2-$\sigma$ agreement for the matter density parameter, being $\Omega_m = 0.403 ^{+0.022}_{-0.024}$ in Bargiacchi results, while here we obtain $\Omega_m = 0.478 \pm 0.041$ (result in Tab.\ref{tablefits}. At the same time for a flat CPL model, we otain an agreement in 1-$\sigma$ ($\Omega_m = 0.447 \pm 0.026$ against $\Omega_m = 0.494 \pm 0.027$).}

It is important now to stress that the QSO sample needs to be calibrated with SNe Ia in the low-redshift region of the Hubble diagram, as already argued in several previous works \citep[see e.g.][]{rl19,2020A&A...642A.150L, 2022MNRAS.515.1795B}. Indeed, QSOs alone are unable to give constraints on any of the free parameters of cosmological models, neither sampled nor derived. 
In this regards, analyses on the use of QSOs alone have been performed \citep[see e.g.][]{2023ApJS..264...46L,2023ApJ...950...45D}. 
However, QSOs can provide significant cosmological information, since they enable the extension of the Hubble diagram up to $z \sim 7.5$. 
Previous works point out that the Hubble diagram of QSOs completely matches the one of SNe Ia in the common low-redshift range, confirming that QSOs are effectively standard candles and showing that the combined Hubble diagram significantly deviates from the prediction of the flat $\Lambda \mathrm{CDM}$ model only at redshifts $z \geq 1.5$, which roughly corresponds to the knee in the Hubble diagram \citep[see e.g.][]{rl15,rl19,lusso19,2020A&A...642A.150L}.

{Since we fit both probes separately and, when justified, different probes together, we employ both individual likelihoods and joint likelihoods. Specifically, joint likelihoods are defined by multiplying the single likelihoods of the combined probes, or equivalently summing the logarithms of the separate likelihoods. This way we avoid introducing any additional normalization and/or factor to weight each probe differently. Indeed, by definition of the single likelihoods, each likelihood is already normalized and thus also the joint likelihood and furthermore each probe is automatically weighted in the joint likelihood according to the number of sources in the data set and its power in constraining the cosmological parameters of the model under study. For the sake of clarity, we here write as an example the likelihoods used separately for QSOs and SNe Ia and their corresponding joint likelihood, employed for the Pth+QSO data set. These likelihoods are defined as follows. Concerning QSOs:
    \begin{equation}
\text{ln}(\mathcal{L})_{\text{QSOs}} = -\frac{1}{2} \sum_{i=1}^{N} \left[ \frac{(\mathrm{log_{10}}L_{X,i}-\mathrm{log_{10}}L_{X,th,i})^{2}}{\hat{s}^{2}_{i}} + \text{ln}(\hat{s}^{2}_{i})\right]
\end{equation}
where $\mathrm{log_{10}}L_{X,i}$ and $\mathrm{log_{10}}L_{X,th,i}$ are the measured and theoretical logarithmic X-ray luminosities, respectively, and $\hat{s}^{2}_{i} = \sigma ^{2}_{\mathrm{log_{10}}L_{X,i}} + \gamma^{2} \sigma ^{2}_{\mathrm{log_{10}}L_{UV,i}} + \delta^{2}$.
While for SNe Ia:
\begin{equation}
\text{ln}(\mathcal{L})_{\mathrm{SNe Ia}} = -\frac{1}{2} \Bigg[\left(\boldsymbol{\mu_{obs, \mathrm{SNe \, Ia}}}-\boldsymbol{\mu_{th}}\right)^{T} \, \textit{C}^{-1} \, \left(\boldsymbol{\mu_{obs, \mathrm{SNe \, Ia}}}-\boldsymbol{\mu_{th}}\right)\Bigg]
\end{equation}
where $\boldsymbol{\mu_{obs, \mathrm{SNe \, Ia}}}$ is the distance modulus measured, $\textit{C}$ is the associated covariance matrix that includes both statistical and systematic uncertainties on the measured distance moduli provided by \textit{Pantheon} release, and $\boldsymbol{\mu_{th}}$ is the distance modulus predicted by the cosmological model assumed, yet depending both on the free parameters of the model and the redshift.
Hence, the joint likelihood reads as:
\begin{equation}
\text{ln}(\mathcal{L})_{\mathrm{QSOs+SNe Ia}} =\text{ln}(\mathcal{L})_{\mathrm{QSOs}} +  \text{ln}(\mathcal{L})_{\mathrm{SNe Ia}}. 
\end{equation}}

\subsection{Dark Energy Survey}
\label{sec_DES}
We use Dark Energy Survey Year-One (DES-1Y) results that 
combine configuration-space two-point statistics from three
different cosmological probes: cosmic shear, galaxy–galaxy lensing, and galaxy clustering, using data
from the first year of DES observations with 1321 square degrees of imaging data \citep{DES:2017myr}. We use the Cobaya implementation of the DES likelihood with covariance matrix, power spectra measurements, and nuisance parameters in agreement with \citep{DES:2017tss}, \citep{,DES:2017qwj}, and \citep{DES:2017myr}. 
We refer to the full combined (3×2pt) sample data set as ``DES".

\subsection{Baryonic Acoustic Oscillations}
\label{sec:BAO}

We include the BAO data from the survey 6dFGS \citep{2011MNRAS.416.3017B}, SDSS DR7 \citep[MGS,][]{2015MNRAS.449..835R}, BOSS DR12 \citep{2017MNRAS.470.2617A} and eBOSS \citep{2018MNRAS.473.4773A}. The first three consist of five BAO measurements between redshift $0.106$ and $0.61$, which represent the most common data set used in recent studies as in \citep{planck2018}, while the last one refers to the measurement at $z=1.52$ from DR14 release. We note here that we do not consider Ly$\alpha$ BAO determinations as they are more complicated and require additional assumptions (e.g. universality of QSO continuum spectra, modeling of spectral lines...) than galaxy BAO measurements, as explained in Section 5.1 of \citep{planck2018}. 

The data set considered in this work is completely consistent with the one used in \citep{2022MNRAS.515.1795B}, but, as already explained in Sect. \ref{sec:qso} for the Pth+QSO data set, also the processing of BAO sample presents some differences, that imply the non-equivalence of the results. 
Specifically, in the present analysis we use Boltzmann codes which calculate the BAO-sensitive parameters for each model considered, while 
the analysis carried out in \citep{ 2022MNRAS.515.1795B} adds a marginalisation over a calibration nuisance parameter. 
Nevertheless, as for the case of Pth+QSO, the BAO analyses of the two works provide completely consistent results within uncertainties.

\section{Cosmological models}
\label{sec:theory}

We now present the cosmological models that we investigate in this work. 
We start with the standard cosmological model, then we extend it by considering a EoS of the cosmological constant, $w$, not fixed to $w= -1$ but free to vary. After that, we considered a scale-dependent EoS, $w(z)$, in its simplest form known as the Chevallier-Polarski-Linder (CPL) parameterisation \citep{Chevallier:2000qy,Linder:2002et}. Finally, we consider a model that deviates from the standard model by considering instead an interaction between the dark components, both in the background and perturbation terms. Specifically, we analyse the following models:

\begin{itemize}
    \item\textbf{$\boldsymbol{\Lambda}$CDM ($\boldsymbol{o\Lambda}$CDM):} The standard cosmological model with zero and non-zero spatial curvature density, $\Omega_{k}$, as a free parameter, which we denote as ``$\Lambda$CDM'' and ``$o\Lambda$CDM'', respectively. 
    
    \item\textbf{$\boldsymbol{w}$CDM ($\boldsymbol{ow}$CDM):} A simple one-parameter extension of the vanilla $\Lambda$CDM model, where $\Lambda$ is no longer a cosmological constant with EoS $w=-1$, but a dark component with EoS $w$ free to vary but still constant with redshift. Also in this case, we explore the model with spatial curvature as a free parameter. Hereafter, the flat and non-flat models are denoted by ``$w$CDM'' and ``$ow$CDM'', respectively;
    
    \item\textbf{CPL:} A two-parameter extension of the standard cosmological model, where the DE component is described by a scale-dependent EoS given by $w\left(a\right)=w_0+w_{a}\left(1-a\right)$, where $a$ is the scale factor defined as $a= 1/ (z+1)$. 
    
    \item\textbf{iDE:} A model that circumvents the problem of the cosmological constant by considering an unified dark sector, which behaves like a decomposed generalized Chaplygin gas (GCG) model with EoS given by 
\begin{equation}
\label{eq:w_ide}
p_{GCG}=-\frac{A}{\rho_{GCG}^{\alpha}},
\end{equation}
where $A$ is a positive constant {depending $\Omega_m$ and $\alpha$, $p_{GCG}$ and $\rho_{GCG}$ are the pressure and the energy density of the decomposed generalized Chaplygin gas, and $\alpha$ is the interaction parameter between the dark components \citep{Fabris:2001tm,Bento:2002ps, Sandvik:2002jz, Bento:2003dj}. }
The GCG interpolates between a CDM, dominating at early times, and a DE component, dominating at late times. The growth rate is suppressed for $\alpha < 0$ due to the homogeneous creation of dark particles from the expanding vacuum, and enhanced when $\alpha > 0$, which implies a clustering of a dynamical DE. In this context, the background evolution can be written as
\begin{equation}\label{eq:Hz_ide}
\frac{H(z)}{H_0} = \sqrt{\left[ (1-\Omega_{m}) + \Omega_{m} (1+z)^{3(1+\alpha)} \right]^{\frac{1}{(1+\alpha)}} + \Omega_{r} (1+z)^4}
\end{equation}
 where $\Omega_{m}$ and $\Omega_{r}$ are the total density of matter and radiation components, respectively, and $(1-\Omega_{m})$ refers to the dark energy density in the flat geometry assumption.  
 In such a model, the $\Lambda$CDM model is recovered for $\alpha = 0$ (for further information we refer to \citep{  vomMarttens:2017cuz,vonMarttens:2018iav,Fabris:2001tm,Alcaniz:2002yt,Bento:2002ps,Bilic:2001cg,Dev:2003cx,Borges:2013bya,Carneiro:2014jza,Wang:2013qy,Bento:2004uh,Carneiro:2017evm,Aurich:2017lck,Ferreira:2017yby,Benetti:2019lxu,Salzano:2021zxk,Benetti:2021div,vonMarttens:2019ixw,Zhu:2004aq} and references within).
\end{itemize}

\section{Method and Results}
\label{sec:Cosmological results}

In our analysis, we first aim to test the physical consistency among the individual probes, which is a crucial point to validate a joint analysis, as detailed in \citep{2022MNRAS.515.1795B}. Indeed, the parameter space of the model analysed with the probes tested individually must show a shared volume in order to combine such data in joint analyses, that is, analyses performed with different data sets must give compatible results. Otherwise, combining incompatible probes would constrain a parameter space that is only the geometric combination of the two, leading to an analysis with no physical meaning \citep{Efstathiou:2020wem,Gonzalez:2021ojp,vagnozzi:2020dfn,Moresco:2022phi,Wang:1999fa}. Hence, we perform a parameter estimation study with each data set presented in Sect.~\ref{sec:Data_sets} assuming a specific cosmological model, then repeating this procedure for each model considered in our work. To this purpose, we need to select a single set of initial conditions and parameters priors that can be used for all analyses, which means to ensure an exploration of the parametric space as similar as possible for all models and for all data sets, and thus a more robust comparison of the results.
This can prove to be a challenging choice that needs to be carefully evaluated since the results may be affected by the initial conditions and choices. On the one hand, any analysed data set provides information on a specific epoch, and thus it is sensitive to a specific set of parameters while it leaves the others largely unconstrained. On the other hand, our goal is to identify a set of parameters that can be left free for all considered data sets, ensuring the use of consistent assumptions for any model as far as possible. We refer to Appendix \ref{appendix:prior} for a discussion on the impact of the initial assumptions on the analysis results.

As a matter of fact, in our analyses involving only large-scale data, such as SNe Ia, QSOs, BAO, and DES, we choose to work with three free parameters, which are $\Omega_{m}$, $\Omega_{b}$, and $H_0$ using wide and flat priors in the following ranges:
\begin{equation}
\nonumber
    \Omega_{m} = [0.01:0.90] , ~~~~\Omega_{b} = [0.01:0.20]
    ~~~~H_0 = [20:200]
\end{equation}
where $H_0$ is in units of $\mathrm{km \, s^{-1} \, Mpc^{-1}}$.
Also, we assume the big bang nucleosynthesis (BBN) consistency, interpolating the primordial helium abundance value provided by table from PArthENoPE fits\footnote{The primordial abundances of helium and deuterium are calculated as a function of baryon density and the extra relativistic number of species. \href{http://parthenope.na.infn.it/}{PArthENoPE
website}.} and we impose a Gaussian prior on the baryon density parameter motivated by observations of D/H abundance \citep{Cooke:2017cwo}, which is $\Omega_bh^2=0.0222\pm0.0005$. At the same time, we fix the remaining cosmological parameters, namely the spectral tilt, $n_s$, the optical depth, $\tau$, the primordial scalar amplitude, $A_s$, and the curvature density, $\Omega_k$, to the following values, as constrained by \citep{planck2018}:
\begin{align*}
n_s &= 0.965 &
\tau &= 0.055 &
A_s &=  2.21 \times 10^{-9}&
\Omega_k &= 0.
\end{align*}
Indeed, the large scale data of SNe Ia, QSOs, BAO, and DES are only sensitive to the components of background evolution, the $H(z)$ equation, with no information on primordial or re-ionization parameters. 
In addition, we consider a free curvature density in the standard cosmological model ($o \Lambda$CDM) and the $o w$CDM one, in order to test the compatibility of the models under this condition as well. 
Instead, when CMB data are considered, all the vanilla cosmological parameters $\{\Omega_b$, $\Omega_{cdm}$, $n_s$, $A_s$, $\tau$, $\theta \}$ (where $\theta$ is the ratio between the sound horizon and the angular diameter distance at decoupling) are assumed free to vary over a wide, flat range. Therefore, we consider models with more free parameters with respect to models analysed with large-scale data, since we assume that fixing them at defined values, although in agreement with the Planck data, would have introduced unjustifiable bias.

Our analyses are performed using the Einstein-Boltzmann Solver "Cosmic Linear Anisotropy Solving System" CLASS~\citep{Blas:2011rf}, which includes the equations needed to describe the primordial perturbations and draw the background expansion of cosmological models. 
This code is interfaced with MCMC algorithm Cobaya \citep{Torrado:2020dgo}, that we use to explore the parameter space and constrain the parameters of the theory. 
The QSO data used in this work are released alongside the publication of this article\footnote{{The complete repository is available at \url{https://github.com/QSO-Cosmology/QSO-for-Cosmology/}. These data are also suitable for use with CosmoMC~\citep{Lewis:2002ah} and MontePython~\citep{Audren:2012wb}.}}.

We consider several models still implemented in such Einstein-Boltzmann solvers, namely the $w$CDM and the CPL parametrisation of DE EoS. Moreover, we also consider a model of interaction between DE and dark matter built from a Chaplygin gas, which is treated as described in \citep{Benetti:2019lxu,Benetti:2021div} and reference therein.

Before presenting our results, it is important to point out that a correct consistency test cannot be performed by investigating only 2D contours separately, but must take into account all the parameter space simultaneously. Actually, 2D graphs are only projections of spaces onto surfaces, that is, slices of analysis, and such projections can sometimes be misleading. 
Therefore, we will only show a few contour planes to illustrate our main results, but our analysis has extended to a complete study of the whole parametric space of the model under consideration \citep{2022MNRAS.515.1795B,Gonzalez:2021ojp}. 
In this regard, we show in the plots and tables only the results obtained from data sets that proved to be compatible, i.e. are in agreement within 3\,$\sigma$ statistical level, as pointed in \citep{Gonzalez:2021ojp}. Then, we consider \textit{in agreement} measurement up to 2 $\sigma$ compatible, while we report a \textit{difference} when the analysis meets at 3 $\sigma$.

In particular, we show the results assuming a flat spatial curvature in Table \ref{tablefits}, while the results obtained by considering the curvature as a free parameter are provided in Table \ref{tab:+Omk}. Below, we detail the results of our analyses.

\begin{figure*}[t!]
\centering
{\includegraphics[width=0.32\linewidth]{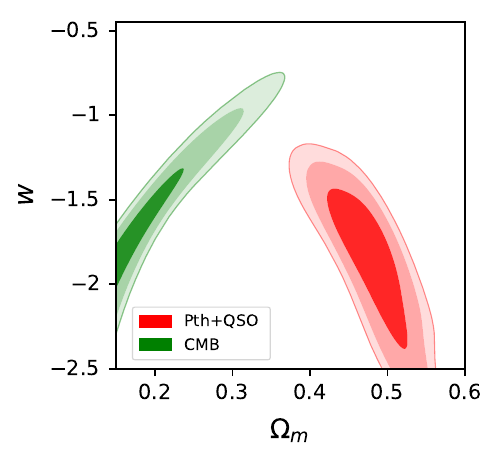}}
{\includegraphics[width=0.30\linewidth]{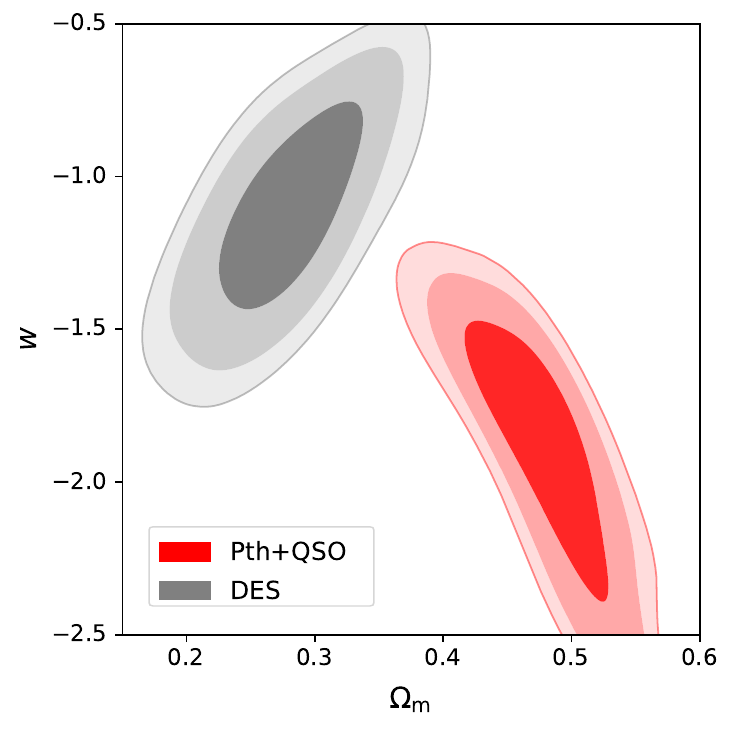}}
{\includegraphics[width=0.32\linewidth]{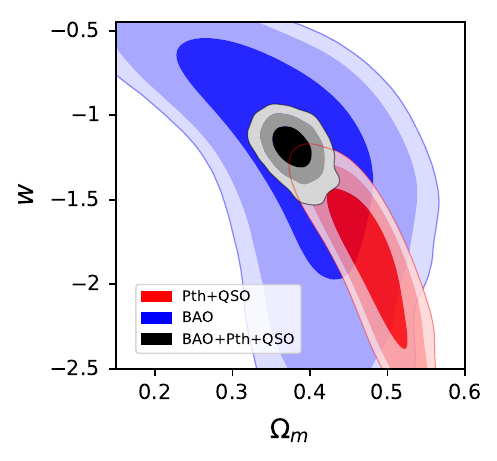}}
    \caption{Bi-dimensional contour plot ($\Omega_{m}$, $w$) at 1, 2 and 3\,$\sigma$ confidence levels obtained from each data set in the legend for the flat $w \mathrm{CDM}$ model analysis.}
    \label{fig:wcdm}
\end{figure*}

\subsection{$\Lambda$CDM}
We start from the investigation of the standard cosmological model, which relies on the cosmological constant $\Lambda$ as the DE component, specifically a fluid that exerts a negative and repulsive pressure, with EoS $w=-1$. Considering a flat geometry of the Universe (see Table \ref{tablefits}), we note an overall agreement between CMB and DES constraints within 1\,$\sigma$, while the values of $H_0$ parameter meet at 2\,$\sigma$ for CMB and BAO. 
Also, in both the Hubble and the matter density values a 2\,$\sigma$ agreement is reached between CMB and Pth+QSO. 
We note a preference of the late-time data, i.e Pth+QSO and BAO, for a higher amount of matter density than that estimated by the CMB, and this effect is in agreement with results of previous works in the literature \citep{2022PhRvD.106d1301O,2023ApJS..264...46L,2023ApJ...951...63D}. 

In general, we note  that for this model the probes are in agreement in at least within 2\,$\sigma$, still showing different slight preferences. It could be read as matter of the standard model, which should include extra physics, implying more degrees of freedom to increase the compatibility between data sets. Indeed, the rigidity of this model, due to the small number of free parameters used in its vanilla description, does not seem to allow it to catch the greater complexity that emerges from the observations. This may hint at the need to extend the simplest cosmological model by considering extra mechanisms and parameters.

The first extension we consider here is a non-flat Universe with the $o \Lambda$CDM model (see Table \ref{tab:+Omk}), where we adopt the prior range for the curvature within [$-$0.5 , 0,5]. We note from our results that Pth+QSO and CMB meet at 3 $\sigma$ on $H_0$, while Pth+QSO are in agreement at $2 \sigma$ on $H_0$ and $\Omega_m$ with DES. The single probes are compatible on the $\Omega_k$ parameter with the exception of Pth+QSO and CMB, which prove to be inconsistent at more than $3 \sigma$. Indeed, Pth+QSO meet DES within $1 \sigma$ and BAO within $2 \sigma$. Thus, only the combination of Pth+QSO+DES and Pth+QSO+BAO are shown in Table \ref{tab:+Omk}. Concerning these combined analyses, the joint data set of Pth+QSO+BAO significantly reduces the uncertainty on $H_0$ and leads to values of $\Omega_m$ and $H_0$ compatible with the ones of the single probes and a negative value of $\Omega_k$ driven by Pth+QSO. Furthermore, Pth+QSO+DES provides a value of $\Omega_m$ towards the one preferred by DES alone, a $H_0$ intermediate between the two data set, and a $\Omega_k$ closer to the flatness compared to the negative values favored by Pth+QSO and DES alone. 

Noteworthy, Pth+QSO constrain matter density and $H_0$ values in agreement in 1 $\sigma$ with the flat case, while the CMB prefers higher values, recovering the flat universe in 2$\sigma$, as also shows in \citep{Planck:2018lbu,2020NatAs...4..196D,2022MNRAS.515.1795B}. 

\subsection{$w$CDM}

The simplest extension of the vanilla $\Lambda$CDM assumption is to consider the DE EoS constant with the scale, but with a value that can be different from $w=-1$. As extensively analysed, the CMB data from the latest Planck release constrains a value of $w$ other than $-$1 in 1\,$\sigma$ \citep{planck2018}, showing a preference for non-standard values. Instead, when the CMB is combined with other data, such as BAO, a value fully compatible with $-$1 is recovered. Therefore, it is reasonable to leave this parameter free and explore precisely the $w$CDM model as the first test of any alternative analysis; hence, instead of $\Lambda$ with $w=-1$ we consider the EoS as a free parameter of the theory, within a flat prior of [$-$3,0]. This changes the background equations and must also be taken into account at the perturbation level.
Below, we report the results of our analysis in two specific situations: considering a flat Universe and relaxing this assumption.

\subsubsection{Flat $w$CDM}

In the case of a flat Universe, the constraints obtained with the Pth+QSO data set (red contours in Fig. \ref{fig:wcdm}) are in tension at more than $3 \sigma$ with CMB (green contours) and DES (gray contours) on the $\Omega_m$ parameter, while they are compatible with the constraints obtained from BAO (blue contours) within 1\,$\sigma$ on all free parameters. 
We stress here that the analyses are conducted strictly under the same initial choices, with the same number of free parameters. This is very important for robust comparison of results and a correct interpretation of the compatibility of the data, which may be affected by the initial choices as shown in Appendix \ref{appendix:prior}.  
The combined BAO+Pth+QSO analysis is placed as a geometric intersection between the analysis of two data sets alone and follows, in terms of the anti-correlation among the parameters, the Pth+QSO analysis, indicating that this sample plays the leading role in constraining the ($w-\Omega_m$) plane. Our results are compatible with \citep{2022MNRAS.515.1795B}, the small differences being attributable to what is already described in Sect. \ref{sec:BAO}. Table \ref{tablefits} also shows the compatibility of Pth+QSO data set with BAO. 

As already discussed, in this paper we adopt the strategy of considering not statistically justified to combine probes that independently do not show compatibility within 3 $\sigma$, as is in the case of the CMB. 
This latter alone manifests a weak power on constraining $H_0$ and $w$, so leaving the EoS free produces a significant deterioration in parameter constraint. 
Given the incompatibility in the $\Omega_m$ estimate between CMB and Pth+QSO, as well as DES and Pth+QSO, it is not statistically and physically correct to explore the combined CMB+Pth+QSO and DES+Pth+QSO data set. Such a joint analysis is thus not indicated in Table \ref{tablefits}, nor shown in Figure \ref{fig:wcdm}. 

These results reveal that a one-parameter extension in the DE sector, compared with the standard cosmological model, preserves compatibility between late-time data but proves unsuitable to describe both CMB and Pth+QSO observations, or even a combination of DES and Pth+QSO.

\subsubsection{$o w$CDM}

Relaxing the flatness assumption, with $\Omega_k$ prior [-0.5 , 0.5], the Pth+QSO data set prefers large total matter density values and high and negative curvature densities (see Table \ref{tab:+Omk}), with $0.397<\Omega_m<0.663$ and $\Omega_k < -0.395$ at $99\%$ confidence level. While Pth+QSO provide only an upper limit of the curvature density, BAO, DES, and CMB give narrow constraints on the curvature, as reported in Table \ref{tab:+Omk}.  Such a constraints are in tension with Pth+QSO upper limit at more than 3\,$\sigma$, and therefore the combined analyses are not reliable, as already stated in \citep{2022MNRAS.515.1795B} for Pth+QSO and BAO.
 
Let us deserve further comment for these analysis. In this case, we see that Pth+QSO data set is unable to jointly constrain the matter density and curvature, due to the degeneracy of these two parameters and the low sensitivity of this probe to $\Omega_k$. The analyses force $\Omega_k$ to the lower values allowed by the adopted prior, and, at the same time, to high values of $\Omega_m$. Extending the $\Omega_k$ prior ranges would shift the values further. Pth+QSO analyses are therefore \textit{unreliable}, and their compatibility (or non-compatibility) with other probes merely driven by the chosen priors.

This allows us also to make some consideration on the model building. On the one hand, it is known that adding physics, parameterised with one more parameter, allows more degrees of freedom to describe the data. On the other hand, the new physics should include a non-degenerate mechanism already described by the base model. Indeed, trying to describe an effect (in this case, the current expansion of the Universe) with the use of two or more parameters could only lead to not being able to constrain either one well. 
Indeed, even where the data are curvature-sensitive, such as the CMB, we notice an increase of the uncertainties in both the $o \Lambda$CDM and $o w$CDM cases compared with the flat case. Choosing the data set in the analysis sensitive to the particular physics one wants to constrain is certainly essential, but the data (or the combination of several samples) should break the degeneracy between parameters, and when they parameterize the same physics, it can prove a strenuous task. 

\begin{figure}[t!]
\centering
{\includegraphics[width=0.36\linewidth]{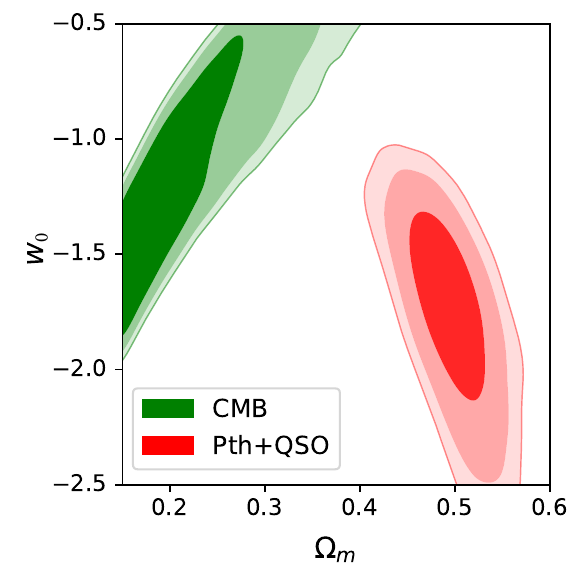}}
{\includegraphics[width=0.36\linewidth]{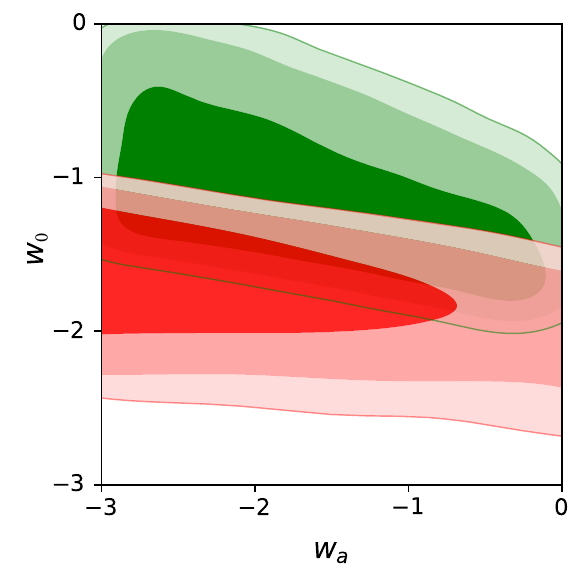}}
{\includegraphics[width=0.36\linewidth]{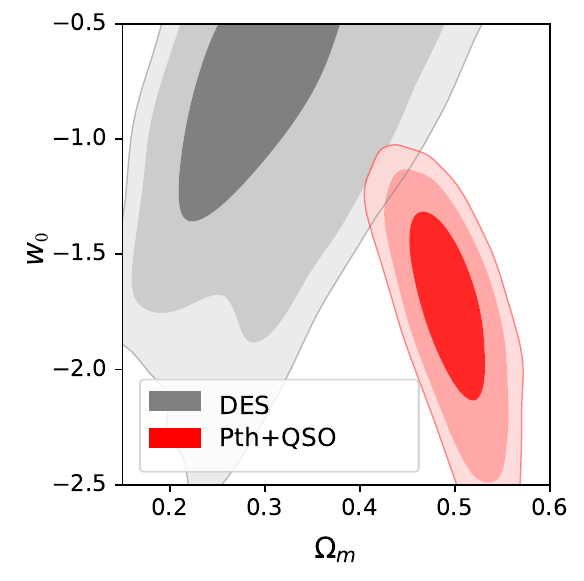}}
{\includegraphics[width=0.36\linewidth]{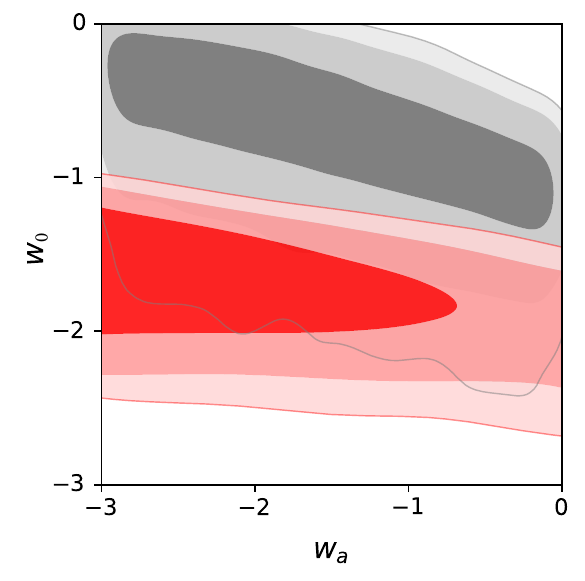}}
{\includegraphics[width=0.36\linewidth]{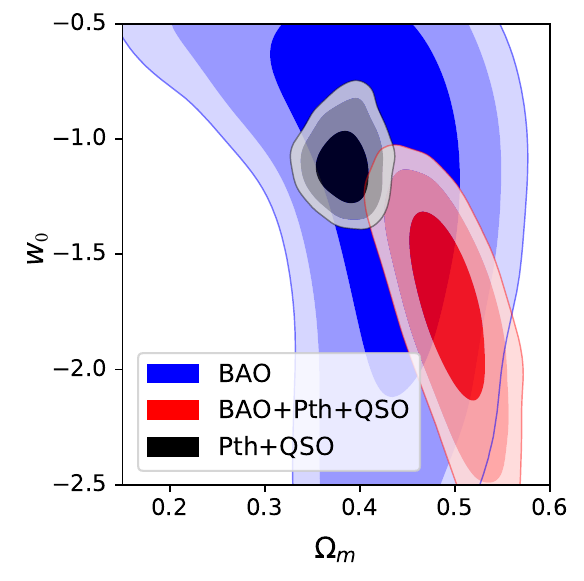}}
{\includegraphics[width=0.36\linewidth]{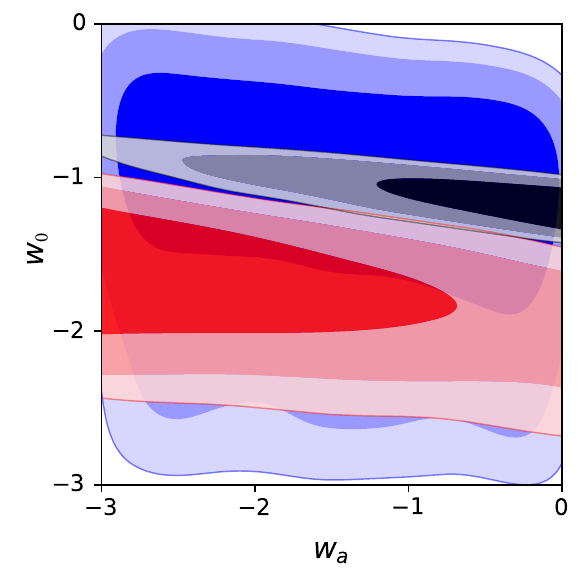}}
\caption{Bi-dimensional contour plot at 1, 2 and 3\,$\sigma$ confidence levels obtained from each data set in the legend for the flat CPL model analysis.}
    \label{fig:cpl}
\end{figure}
\subsection{Flat CPL}

The CPL parameterisation \citep{Chevallier:2000qy, Linder:2002et} is the easiest model to consider a dynamic evolution of DE. This model considers two further parameters of the theory, one which determines the value of the DE EoS today ($w_0$), and one which parameterises its evolution with the scale ($w_a$). Being parameters describing late physics mechanism, the CMB is proved to be not very sensitive to these degrees of freedom, and more decisive in constraining their values are the measurements of BAO, Redshift Space Distortion (RSD), SNe Ia (see the full analysis of \citep{planck2018}) and Fast Radio Bursts \citep[FRBs;][]{Zhao:2020ole}. These data constrain a $w_a$ value consistent with zero, although slightly negative values are preferred. 
Current data of CMB+BAO+SNe Ia constrain $w_0= -0.957 \pm 0.080$ and $w_a = -0.29^{+0.32}_{-0.26}$ \citep{planck2018}, while Pantheon+ SNe Ia sample with SH$0$ES Cepheids distance provides $w_0=-1.81^{+1.71}_{-0.60}$ and $w_a = -0.4^{+1.0}_{-1.8}$ \citep{Brout:2022vxf}.

We find that no data set imposes constraints on $w_a$. As shown by the plots on the right side of Fig.~\ref{fig:cpl},  $w_0$ demonstrates no significant correlation with $w_a$ across all data sets. 

Since $w_a$ is unconstrained, the CPL model restricts a parametric space very similar to that of the $w$CDM model, and the addition of a DE scale dependence does not lead compatibility between CMB and Pth+QSO (top panel of Fig.~\ref{fig:cpl}) and DES and Pth+QSO (middle panel of Fig.~\ref{fig:cpl}), leaving instead BAO consistent with Pth+QSO. 
Also, CMB (and DES) do not constrain $w_0$ parameter, only imposing an upper (lower) limit on the DE EoS.  
\begin{figure}[t!]
\centering
{\includegraphics[width=0.51\linewidth]{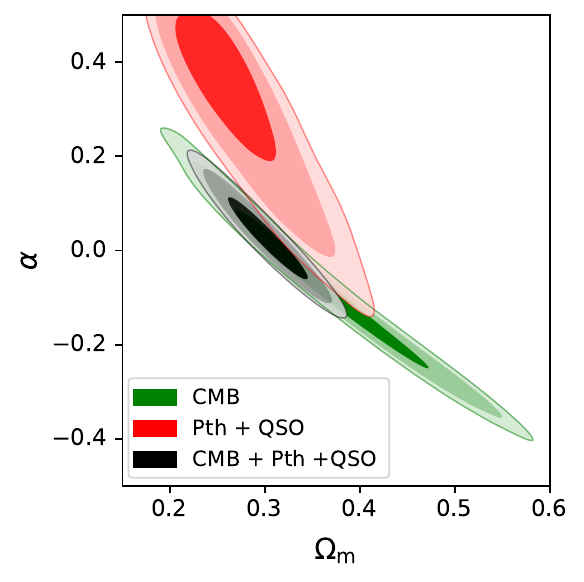}}
{\includegraphics[width=0.51\linewidth]{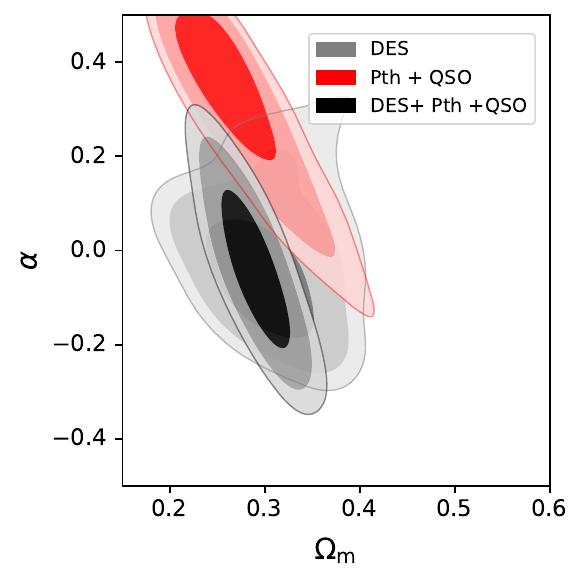}}
{\includegraphics[width=0.51\linewidth]{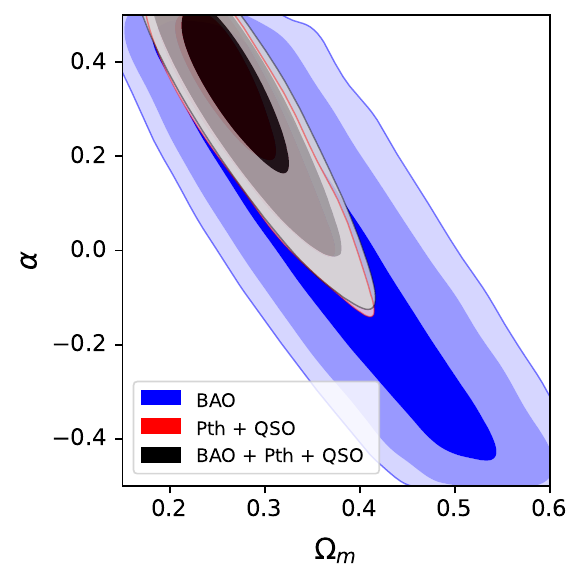}}
\caption{Bi-dimensional contour plot at 1, 2 and 3\,$\sigma$ confidence levels obtained from each data set in the legend for the flat iDE model analysis.}
    \label{fig:ide}
\end{figure}
{ In order to further explore the behavior of the CPL parametrization, we have considered a reduced scenario in which $w_0$ is fixed to the cosmological constant value, and only the parameter $w_a$ is left free to vary. This test isolates the impact of a possible time variation in the dark energy equation of state, independently of its present-day value. As expected, fixing $w_0$ helps reduce the allowed parameter space for $w_a$, with significantly tighter constraints obtained from the same datasets. However, the general degeneracy structure observed in the full CPL model remains qualitatively unchanged.} 
{Conversely, we note that fixing $w_a$ while allowing $w_0$ to vary essentially corresponds to a $w$CDM model, which is already discussed in detail in the main analysis. Moreover, assuming non-zero fixed values of $w_a$ would not result in meaningful improvements, as it would mostly shift the inferred central value of $w_0$ without altering the overall trends.}
\subsection{Flat iDE}

We consider here a GCG model that allows for non-gravitational interactions between the dark components via a dissipation mechanism. Such a mechanism can generate either the creation of dark particles from the expanding vacuum or an increasing effect of dynamical DE. The dissipation is characterized by the parameter $\alpha$, which behaves at the background level like cold matter in the early times and tends to be a cosmological constant in the asymptotic future. 
A value of $\alpha$ less than zero would indicate the creation of DM from DE, whilst a positive value the opposite. Previous works have shown that late-universe data, such as LSS (2dFGRS) and SNe Ia (JLA), prefer negative values of $\alpha$ \citep{Pigozzo:2015swa}, while the inclusion of CMB calls for values more compatible with zero \citep{vomMarttens:2017cuz, Aurich:2017lck, Benetti:2019lxu, Salzano:2021zxk,Benetti:2021div,Borges:2023xwx,vonMarttens:2019wsc}, and SNe+QSO favour positive values \citep{Zheng:2022vhj,2023A&A...678A..13G}.

In our analyses, we explore this model using a wide, flat prior in the $\alpha$ range of [$-$0.5 , 0.5]. Our results show that this model allows reconciling CMB, BAO, and DES with Pth+QSO, as shown in Fig. \ref{fig:ide} and Table \ref{tablefits}. 
Interestingly, we note that the prediction of the standard model of a larger amount of $\Omega_m$ using the Pth+QSO sample than predicted by the CMB is solved by introducing the GCG dissipation mechanism. Indeed, the high positive values of $\alpha$ means dissipation of vacuum energy into dark particles, and thus captures the extra amount of CDM predicted by the QSOs, thus reconciling the cosmological parameter estimates. In general, the dissipation values for each analysis are always compatible with zero in 1\,$\sigma$ and thus they recover the $\Lambda$CDM model, with the exception of Pth+QSO, which instead recovers the standard model in 3\,$\sigma$.  
It is also worth noting that this model produces a widening of the error bars of the cosmological parameters using CMB data, compared with the $\Lambda$CDM model, allowing it to arrive in 3\,$\sigma$ at the maximum value of $H_0 = 70.92 \, \mathrm{km \, s^{-1}\, Mpc^{-1}}$.

Combining data from Pth+QSO with the other probes makes $\alpha$ compatible with the standard model, with the exception of the Pth+QSO+BAO data set, which instead shows a clear preference for $H_0$ values compatible with direct observations of SNe Ia and a non-zero dissipation mechanism between the dark components. This is because BAOs are not proved to be sensitive to the dissipation parameter (see bottom panel of Fig. \ref{fig:ide}) so the Pth+QSO data set is driving the $\alpha$ parameter constraint. 
The combined analyses shown in Fig. \ref{fig:ide} lie as the geometric intersection of the contour levels of the two data sets analysed independently only for the combination with BAOs. As for DES, the combined analysis lies on DES and acquires the parameter correlation of Pth+QSO. As for CMB, the combined analysis lies on CMB but $\alpha$ is restricted in the negative values as shown by Pth+QSO.

\begin{table*}
\caption{Mean values and 1\,$\sigma$ uncertainties for the cosmological free parameters in each model with zero curvature and data set considered in this work. When upper limit values are indicated, these refer to the value in 2\,$\sigma$. Furthermore, when the probability distribution is not Gaussian, the value is indicated with the two distinct limits of the 1\,$\sigma$ error. }
\label{tablefits}
\centering 
\setlength{\tabcolsep}{4pt}
\renewcommand{\arraystretch}{1.5}
\scalebox{0.7}{
\begin{tabular}{ c c c c c c c } 
\hline\hline
Model & Data set & $\Omega_{m}$ &  $H_0 \, (\mathrm{km \, s^{-1} \, Mpc^{-1}})$ & $w_0$ & $w_{a}$ & $\alpha$\\
\hline
Flat $\Lambda \mathrm{CDM}$ 
& {Pth+QSO} & $ 0.357 \pm 0.025 $ & $ 71.79 \pm 1.74 $ &-1 &-&- \\
& CMB & $0.316 \pm  0.007$&  $67.34 \pm 0.54$ &-1 &-&- \\
& BAO & $0.371_{-0.053}^{+0.044}$&  $72.06^{+3.53}_{-2.68}$ &-1 &-&- \\
& DES & $0.300 \pm 0.049$&  $66.76 \pm 7.79$ &-1 &-&- \\
& {Pth+QSO+CMB} & $0.316 \pm 0.008$&  $67.33 \pm 0.60$ &-1 &-&- \\
& {Pth+QSO+BAO} & $0.358 \pm 0.024$&  $71.14 \pm 2.30$ &-1 &-&- \\
& {Pth+QSO+DES} & $0.285 \pm 0.014$&  $69.44 \pm 2.35$ &-1 &-&- \\
Flat $w\mathrm{CDM}$ 
&  {Pth+QSO} & $ 0.478 \pm 0.041 $ & $ 71.70 \pm 4.97$ & $ -1.87 \pm 0.28 $&-&- \\ 
& CMB & $0.130 ^{+0.018}_{-0.067}$ &  $111.70 ^{+28.43}_{-20.70}$ & $< -1.24$ &-&- \\ 
& BAO & $0.377 ^{+0.081}_{-0.053}$&  $78.31^{+12.50}_{-16.54}$ &$-1.26^{+0.62}_{-0.27}$ &-&- \\
& {DES} & $0.281 \pm 0.033$&  $68.53 \pm 4.73$ &$-1.10 \pm 0.21$ &-&- \\ 
& {Pth+QSO+BAO} & $0.378 \pm 0.017$&  $75.33 \pm 1.74$ &$-1.19 \pm 0.08$ &-&- \\ 
Flat CPL
& {Pth+QSO} & $ 0.494 \pm 0.027$ & $ 73.37 \pm 1.76 $ & $ -1.74 \pm 0.28 $ & unconstrained &- \\
& CMB & $0.219 ^{+0.028}_{-0.091}$ &  $82.92 \pm 10.71$ & $< - 0.28$ & unconstrained & - \\ 
& BAO & $0.414 ^{+0.067}_{-0.045}$&  $82.87 ^{+13.98}_{-17.12}$ & $-1.18 ^{+0.76}_{-0.38}$ &unconstrained &- \\
& DES & $0.306 ^{+0.053}_{-0.069}$&  $65.74 \pm 8.57$ & $> -1.48$ &unconstrained&- \\ %
& {Pth+QSO+BAO} & $0.383 \pm 0.018$&  $75.54 \pm 1.73$ & $-1.11 \pm 0.10$ & $>-1.96$ &- \\
& {Pth+QSO+DES} & $0.282 \pm 0.009$&  $70.56 \pm 1.61$ & $-0.90 \pm 0.05$ & $> - 0.54$ &- \\ %
Flat iDE 
& Pth+QSO & $0.264 ^{+0.018}_{-0.025}$ & $ 73.35 \pm 1.82$ &-1 &-& $0.34 ^{+0.15}_{-0.06}$ \\
& CMB & $0.364 ^{+0.057}_{-0.081}$ & $ 65.61 ^{+2.73}_{-2.15}$ &-1 &-& $-0.08 \pm 0.11 $\\ %
& BAO & $0.359 ^{+0.045}_{-0.060}$ & $ 71.25 \pm 3.83$ &-1 &-& $0.04 ^{+0.44}_{-0.20}$ \\ %
& DES & $0.303 ^{+0.032}_{-0.038}$ & $68.77\pm {5.55}$ &-1 &-& $-0.06 \pm 0.09$ \\ %
& Pth+QSO+CMB & $0.302\pm 0.024$&  $67.79 \pm 0.91$ &-1 &-& $0.03 \pm 0.05$ \\ %
& Pth+QSO+BAO & $0.271 ^{+0.031}_{-0.040}$ & $ 73.98 ^{+2.03}_{-1.99}$ &-1 &-& $0.34 ^{+0.15}_{-0.06}$ \\ %
& Pth+QSO+DES & $0.289 \pm 0.021 $ & $ 71.17 \pm 3.21 $ &-1 &-& $-0.04 \pm 0.10 $ \\ %
 \hline
\end{tabular}}
\end{table*}

\section{Summary \& Conclusions}
\label{Conclusions}

The current availability of increasingly accurate data revealed observational inconsistencies with the predictions of the standard cosmological model. While it might indicate the need to study possible extensions of the standard theory, it seems crucial to search for observations on scales where they are currently lacking. QSOs thus prove to be key objects for extending the Hubble diagram to redshifts greater than those of SNe, allowing exploration up to $z=7.64$.

In previous work of \citep{2022MNRAS.515.1795B}, the QSOs are first analyzed to test DE extension models and compared with BAO data. In this work, we investigated the standard cosmological model and some of its extensions using QSOs, SNe Ia, BAO, DES, and CMB data as cosmological probes. In particular, we included for the first time QSOs in the MCMC algorithms CosmoMC, Montepython, and Cobaya by applying a method originally developed by our group. 
As cosmological models, we considered the flat and non-flat $\Lambda$CDM and $w$CDM models, the CPL parameterisation, and a iDE scenario within a decomposed Chaplygin gas model. All these models are fitted with each individual probe and then the consistency of these probes is checked in the multi-dimensional space of free parameters to explore the possibility of a combined analysis. Indeed, as already pointed out in \citep{Gonzalez:2021ojp} and \citep{2022MNRAS.515.1795B}, the physical compatibility of the constraints obtained from each data set is a necessary requirement to combine different probes. 

From our analysis, summarized by Tables \ref{tablefits} and \ref{tab:+Omk} and Figs.\ref{fig:wcdm}, \ref{fig:cpl}, and \ref{fig:ide}, we show that:

\begin{itemize}
    \item  the $\Lambda$CDM model is able to agree Pth+QSO data with the other probes considered, constraining the free parameters values within 2$\sigma$. 

    \item in the o$\Lambda$CDM model, BAO and DES probes are still compatible with Pth+QSO, while CMB is in tension at more than 3$\sigma$. The Pth+QSO only imposes an upper limit on the density curvature parameter. These results indicate a clear failure of this model to simultaneously describe information at all scales.

    \item extending the DE sector with a constant equation of state not fixed at $w=-1$, does not help to make CMB and DES consistent with Pth+QSO (DES is again found to agree), and it only gets worse if we extend this model by introducing a non-zero curvature. Indeed, none of the probes are in agreement with Pth+QSO in the $ow$CDM model, since the latter turns out to prefer large, negative values of curvature density. 

    \item with the CPL model, the data seems to be not sensitive to the parameter that governs the variation of the equation of state with scale, $w_a$, so the parametric space restricted by this model is essentially the same as the $w$CDM model (where $w_a=0$), with equal or larger error bars depending on the probe considered.

    \item the GCG iDE model here considered seems to be able to reconcile CMB and Pth+QSO by preferring an interaction that generates DM particles from vacuum energy. Also, DES and BAO probes are found to be in agreement with all other data sets.
    
\end{itemize}

Our results point out that where the simplest models of DE extension fail in describing the Universe evolution at all scales, the more complicated model involving DE-DM interactions can succeed. This would therefore show the path to be explored to make several cosmological and astrophysical observations compatible, especially in light of future QSO data that will be available from the Euclid and Athena collaborations \citep{Tutusaus:2022aef,Matt:2019llr}.

\begin{table*}
\caption{Mean values and 1\,$\sigma$ uncertainties for the cosmological free parameters in each model with curvature and data set considered in this work. When upper limit values are indicated, these refer to the value in 2\,$\sigma$. Furthermore, when the probability distribution is not Gaussian, the value is indicated with the two distinct limits of the 1\,$\sigma$ error.}
\label{tab:+Omk}
\centering 
\setlength{\tabcolsep}{4pt}
\renewcommand{\arraystretch}{1.5}
\scalebox{0.8}{
\begin{tabular}{ c c c c c c } 
\hline\hline
Model & Data set & $\Omega_{m}$ &  $H_0  \, (\mathrm{km \, s^{-1} \, Mpc^{-1}})$ & $w$ & $\Omega_k$ \\
\hline
$o \Lambda$CDM 
& {Pth+QSO} & $ 0.376 \pm 0.024$&  $ 72.50 \pm 2.04 $ &-1 & $ <-0.061 $ \\ 
& CMB & $0.351 \pm 0.024$&  $63.61 \pm 2.27$ &-1 & $-0.010 \pm 0.007$ \\
& BAO & $0.362 \pm 0.060$&  $72.04 \pm 7.5$ &-1 & $-0.005 \pm 0.120$ \\
& DES & $0.285 \pm 0.044$&  $67.42 \pm 6.0$ &-1 & $-0.034 \pm 0.068$ \\ 
&{Pth+QSO+BAO} & $0.366 \pm 0.018$&  $73.63 \pm 1.59$ &-1 & $<-0.049$ \\ 
&{Pth+QSO+DES} & $0.280 \pm 0.011$&  $ 70.17 \pm 1.65$ &-1 & $-0.007 \pm 0.033$ \\
$o w$CDM
& {Pth+QSO} & $ 0.536 \pm 0.054$ & $ 73.87 \pm 1.76 $ & $ -1.11 \pm 0.12$ &$  <-0.433 $ \\ 
& CMB & $0.237^{+0.053}_{-0.178}$&  $ 88.27 ^{+18.21}_{-35.54}$ & $<-0.57 $&$-0.007 ^{+0.010}_{-0.001}$ \\
& BAO & $0.376 \pm 0.056$&  $76.41 \pm 8.22$ & $-1.31 \pm 0.38 $ & $0.048 ^{+0.104}_{-0.074}$\\ 
& DES &$0.291 \pm  0.047$&  $67.23 \pm 6.75$ & $-0.92 \pm 0.17 $ & $-0.067 ^{+0.067}_{-0.101}$ \\
 \hline
\end{tabular}}
\end{table*}

\section*{Acknowledgments}
GB, MB, and SC acknowledge Istituto Nazionale di Fisica Nucleare (INFN), sezione di Napoli e Laboratori Nazionali di Frascati (LNF), \textit{iniziative specifiche} QGSKY and MOONLIGHT-2. 
MS acknowledges financial support from the Italian Ministry for University and Research, through the grant PNRR-M4C2-I1.1-PRIN 2022-PE9-SEAWIND: Super-Eddington Accretion: Wind, INflow and Disk-F53D23001250006-NextGenerationEU.\\
This paper is based upon work from COST Action CA21136 {\it Addressing observational tensions in cosmology with systematics and fundamental physics} (CosmoVerse) supported by COST (European Cooperation in Science and Technology).

\begin{appendix}

\section{Role of analysis assumptions}
\label{appendix:prior}

Recently, more and more importance has been given to the statistical validity of combining data sets in joint analyses and properly comparing models. Similarly attention should be deserved to the initial conditions of the analysis, as the setting of parameters priors and boundary choices significantly influence the results. 
While it is true that the choice of fixing certain parameters while allowing others to vary is often taken arbitrarily in a given model, attention should be paid to select the parameters of the theory that are sensitive to the data sets analysed, since leaving one of them fixed necessarily means spoiling the results. Analogously, it is also important to choose a sufficiently high prior range so that the parametric space is duly explored.

\begin{figure}[!]
\centering
{\includegraphics[width=0.7\linewidth]{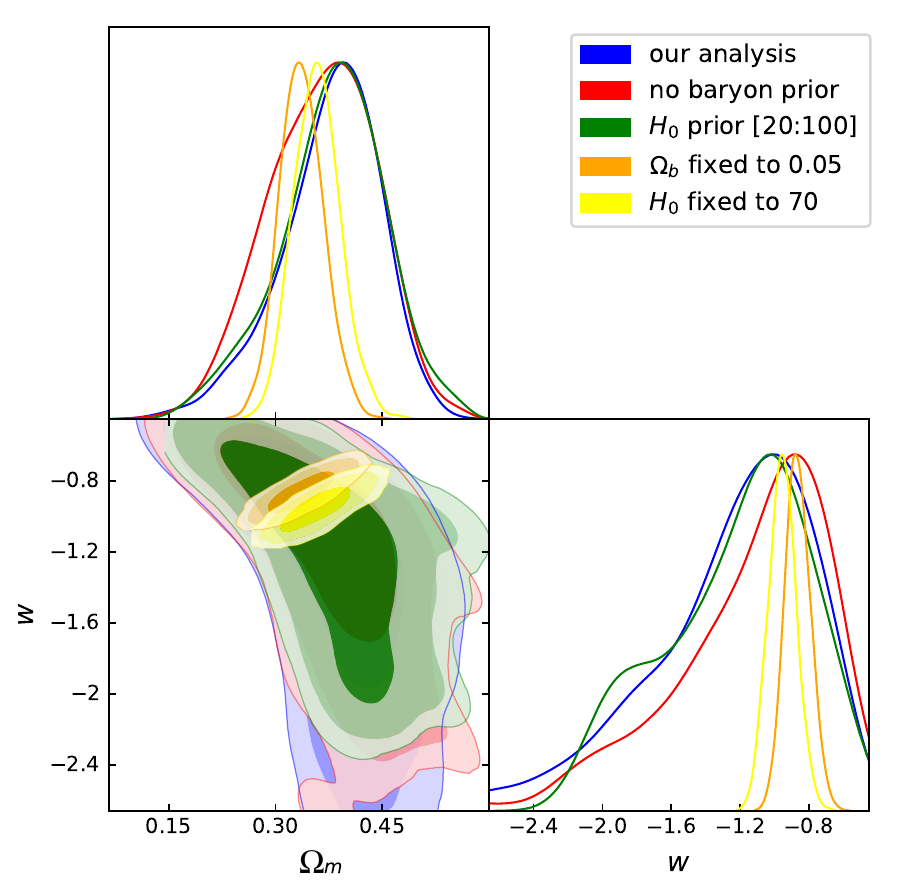}}
    \caption{Analysis of the flat $w \mathrm{CDM}$ model, using BAO data and different assumptions with respect to the ones used in this work, which consist of leaving free parameters $\Omega_{m}$, $\Omega_{b}$, and $H_0$ assuming BBN consistency and D/H abundance prior on $\Omega_{b}h^2$ (this analysis is drawn with blue line)}.
    \label{fig:wcdm_comparison}
\end{figure}

As a didactic example, we would like to show here the significant changes in the parameter constraints for the flat $w$CDM model, when we vary some of the assumptions used in the analyses in this paper\footnote{As explained in Sect. \ref{sec:Cosmological results}, we consider as free parameters of the MCMC analysis $\Omega_{m}$, $\Omega_{b}$, and $H_0$, imposing wide and flat priors as $\Omega_{m} = [0.01:0.60]$ , $\Omega_{b} = [0.01:0.20]$, and $H_0 = [20:200]$. Also, the BBN consistency and a Gaussian prior motivated by observations of D/H abundance \citep{Cooke:2017cwo} are considered.}. For our purposes, we study the analysis with BAO data, introduced in Sect. \ref{sec:BAO}. 
We here stress that the BAO measurements of volume averaged distance, $D_V(z)$, depends on $H(z)$ evolution and the co-moving angular diameter distance, $D_M(z)$, that also depends on the $H_0$ value. It means that considering the parameters of matter densities and the local value of the Hubble constant is essential for an analysis with BAO data using measures of $D_V(z)$ and $D_M(z)$ \citep{Aubourg:2014yra}. In addition, it is important to break the degeneration between baryonic and dark matter by using the prior provided by the primordial abundance of BBN, as well as impose the constraint on baryonic matter that can reproduce the current observations. We draw an analysis with the characteristics used in this paper in blue in Fig. \ref{fig:wcdm_comparison}, compared with the results obtained by changing these initial assumptions. For example, we can notice that reducing the prior of the $H_0$ parameter to a maximum of 100 (green curves), as it is generally employed in cosmological analyses, reduces the parametric space exploration of the EoS parameter $w$ in 2\,$\sigma$ contour levels. This can be significant but not decisive, since large deviations from $w=-1$ are generally not taken into account. Vice versa, fixing $H_0$ to a specific value (i.e. 70 in our case) tightens both the $w$ and total density of matter constraints also changing the correlation among parameters, as shown in the yellow curves, with a considerable impact on the results and their physical interpretation. 
Even neglecting the prior on the baryonic density can slightly change the parameter space (red curves), while fixing the value of the baryonic matter density (orange curves) significantly changes the results with similar impact as the choice of fixing $H_0$.

We can infer from these tests that reducing the parameter space exploration of the current Hubble constant value or the current baryonic density can lead to misleading results both on the compatibility of the data sets and in properly comparing different models, if BAO data are used. Indeed, estimating Bayesian evidence for a given theory implies weighing its MCMC exploration in the space of all its parameters, and one must always consider wide priors to guarantee fair estimates.

\end{appendix}

\bibliographystyle{elsarticle-num} 
\bibliography{bibl}


\end{document}